\documentclass[10pt,twocolumn,twoside]{IEEEtran}

\usepackage{mathbf-abbrevs}
\newcommand{\reals}{{\mathbb R}}
\newcommand{\ints}{{\mathbb Z}}

\newcommand{\term}{\emph}

\renewcommand{\mid}{\; ; \;}

\newcommand{\floor}[1]{{\left\lfloor #1 \right\rfloor}}
\newcommand{\sfloor}[1]{{\lfloor #1 \rfloor}}

\newcommand{\fracpart}[1]{\left\langle #1 \right\rangle}
\newcommand{\sfracpart}[1]{\langle #1 \rangle}

\usepackage[square,comma,numbers,sort&compress]{natbib}
\usepackage{units}
\usepackage{booktabs}
\usepackage{amsmath,amsfonts,amssymb, amsthm, bm} % this is handy for mathematicians and physicists
\usepackage[vlined, linesnumbered, ruled]{algorithm2e}
\usepackage{mathrsfs}
\usepackage{tikz}
\usetikzlibrary{calc}
\usepackage{pgfplots}
\usetikzlibrary{pgfplots.groupplots}

\newtheorem{proposition}{Proposition}

\usepackage{bibspacing}
\title{Basis construction for range estimation by phase unwrapping}
\author{Assad Akhlaq, R.~G.~McKilliam, and R. Subramanian 
\thanks{The authors are at the Institute for Telecommunications Research, The University of South Australia, SA, 5095.  Supported by Australian Research Council Linkage Project LP130100514.}
}

\begin{document}
%\ninept
% 
\maketitle

\begin{abstract}

%The abstract should appear at the top of the left-hand column of text, about
%0.5 inch (12 mm) below the title area and no more than 3.125 inches (80 mm) in
%length.  Leave a 0.5 inch (12 mm) space between the end of the abstract and the
%beginning of the main text.  The abstract should contain about 100 to 150
%words, and should be identical to the abstract text submitted electronically
%along with the paper cover sheet.  All manuscripts must be in English, printed
%in black ink.

We consider the problem of estimating the distance, or range, between two locations by measuring the phase of a sinusoidal signal transmitted between the locations. This method is only capable of unambiguously measuring range within an interval of length equal to the wavelength of the signal. To address this problem signals of multiple different wavelengths can be transmitted.  The range can then be measured within an interval of length equal to the least common multiple of these wavelengths.  Estimation of the range requires solution of a problem from computational number theory called the \emph{closest lattice point} problem.  Algorithms to solve this problem require a \emph{basis} for this lattice.  Constructing a basis is non-trivial and an explicit construction has only been given in the case that the wavelengths can be scaled to pairwise relatively prime integers.  In this paper we present an explicit construction of a basis without this assumption on the wavelengths.  This is important because the accuracy of the range estimator depends upon the wavelengths.  Simulations indicate that significant improvement in accuracy can be achieved by using wavelengths that cannot be scaled to pairwise relatively prime integers.

\end{abstract}
\section{Introduction}\label{sec:intro}

\newcommand{\lcm}{\operatorname{lcm}}

Range (or distance) estimation is an important component of modern technologies such as electronic surveying~\cite{Jacobs_ambiguity_resolution_interferometery_1981, anderson1998surveying} and global positioning~\cite{Teunissen_GPS_LAMBDA_2006,Teunissen_GPS_1995}.  Common methods of range estimation are based upon received signal strength~\cite{Chitte_RSS_Estimation2009, HingCheung_RSSbasedRangeEstimation2012}, time of flight (or time of arrival)~\cite{XinrongLi_TOA_range_estimation2004, Lanzisera_TOA_range_estimation2011}, and phase of arrival~\cite{Jacobs_ambiguity_resolution_interferometery_1981,Towers_frequency_selection_interferometry_2003,Li_distance_est_wrapped_phase}.  This paper focuses on the phase of arrival method which provides the most accurate range estimates in many applications.  Phase of arrival has become the technique of choice in modern high precision surveying and global positioning ~\cite{Odijk-nteger-ambiguity-resolutionPPP, Teunissen_GPS_LAMBDA_2006, Teunissen_GPS_1995}.

A difficulty with phase of arrival is that only the principal component of the phase can be observed.  This limits the range that can be unambiguously estimated.  One approach to address this problem is to utilise signals of multiple different wavelengths and observe the phase at each.   %The range can then be measured within an interval of length equal to the least common multiple of these wavelengths.  
Range estimators from such observations have been studied by numerous authors~\cite{Teunissen_GPS_1995,Hassibi_GPS_1998,Towers_frequency_selection_interferometry_2003,Li_distance_est_wrapped_phase}.  Least squares/maximum likelihood and maximum a posteriori (MAP) estimators of range have been studied by Teunissen~\cite{Teunissen_GPS_1995}, Hassibi and Boyd~\cite{Hassibi_GPS_1998}, and more recently Li~et.~al.~\cite{Li_distance_est_wrapped_phase}.  A key realisation is that least squares and MAP estimators can be computed by solving a problem from computational number theory known as the~\emph{closest lattice point problem}~\cite{Babai1986,Agrell2002}.  Teunissen~\cite{Teunissen_GPS_1995} appears to have been the first to have realised this connection. %Hassibi and Boyd~\cite[Sec. VII]{Hassibi_GPS_1998} consider weighted least squares and MAP estimators

Efficient general purpose algorithms for computing a closest lattice point require a~\emph{basis} for the lattice.  Constructing a basis for the least squares estimator of range is non-trivial.  Based upon the work of Teunissen~\cite{Teunissen_GPS_1995}, and under some assumptions about the distribution of phase errors, Hassibi and Boyd~\cite{Hassibi_GPS_1998} construct of a basis for the MAP estimator.  Their construction does not apply for the least squares estimator.\footnote{The least squares estimator is also the maximum likelihood estimator under the assumptions made by Hassibi and Boyd~\cite{Hassibi_GPS_1998}.  The matrix $G$ in~\cite{Hassibi_GPS_1998} is rank deficient in the least squares and weighted least squares cases and so $G$ is not a valid lattice basis.  In particular, observe that the determinant of $G$~\cite[p.~2948]{Hassibi_GPS_1998} goes to zero as the a priori assumed variance $\sigma_x^2$ goes to infinity.}  This is problematic because the MAP estimator requires sufficiently accurate prior knowledge of the range, whereas the least squares estimator is accurate without this knowledge.  An explicit basis construction for the least squares estimator was recently given by Li~et.~al.~\cite{Li_distance_est_wrapped_phase} under the assumption that the wavelengths can be scaled to pairwise relatively prime integers.  In this paper, we remove the need for this assumption and give an explicit construction in the general case.  This is important because the accuracy of the range estimator depends upon the wavelengths.  Simulations show that a more accurate range estimator can be obtained using wavelengths that are suitable for our basis, but are not suitable for the basis of Li~et.~al.~\cite{Li_distance_est_wrapped_phase}.

The paper is organised as follows.  Section \ref{sec:ls-estimator} presents the system model and defines the least squares range estimator. Section \ref{sec:lattice-theory} introduces some required properties of lattices.  Section~\ref{sec:range-estim-clos} shows how the least squares range estimator is given by computing a closest point in a lattice.  An explicit basis construction for these lattices is described. Simulation results are discussed in Section~\ref{sec:simulation-results} and the paper is concluded by suggesting some directions for future research.

\section{Least squares estimation of range}\label{sec:ls-estimator}

%Consider the problem of estimating the distance between a transmitter and a receiver using phase only measurements. 
Suppose that a transmitter sends a signal $x(t) = e^{2\pi(ft + \phi)}$ with phase $\phi$ and frequency $f$ in Hertz.  The signal is assumed to propagate by line of sight to a receiver resulting in the signal 
\[
y(t) = \alpha x(t - r_0/c) + w(t) = \alpha e^{2\pi(ft + \theta)} + w(t)
\]
where $r_0$ is the distance (or range) in meters between receiver and transmitter, $c$ is the speed at which the signal propagates in meters per second, $\alpha > 0$ is the real valued amplitude of the received signal, $w(t)$ represents noise, $\theta = \phi - r_0/\lambda$ is the phase of the received signal, and $\lambda = c/f$ is the wavelength.  The receiver is assumed to be \emph{synchronised} by which it is meant that the phase $\phi$ and frequency $f$ are known to the receiver.  

Our aim is to estimate $r_0$ from the signal $y(t)$.  To do this we first calculate an estimate $\hat{\theta}$ of the principal component of the phase $\theta$.  In optical ranging applications $\hat{\theta}$ might be given by an interferometer.  In sonar or radio frequency ranging applications $\hat{\theta}$ might be obtained from the complex argument of the demodulated signal $y(t)e^{-2\pi f t}$.  Whatever the method of phase estimation, the range $r_0$ is related to the phase estimate $\hat{\theta}$ by the phase difference
\begin{equation}
Y = \sfracpart{\phi - \hat{\theta}} = \sfracpart{ r_0/\lambda + \Phi }
\end{equation}
where $\Phi$ represents phase noise and $\sfracpart{x} = x - \floor{x + \tfrac{1}{2}}$ where $\floor{x}$ denotes the greatest integer less than or equal to $x$.  %That is, $\fracpart{x}$ is the \emph{centered} fractional part of $x$. %For the purpose of analysis, it is convenient to model the phase noise using the~\emph{wrapped normal distribution}~\cite[p.~50]{Mardia_directional_statistics}\cite[p.~76]{McKilliam2010thesis}\cite[p.~47]{Fisher1993}.  That is, we model the phase noise as $\fracpart{\Phi}$ where $\Phi$ is normally distributed with zero mean. With this assumption $\hat{\theta} = \fracpart{ \theta - \fracpart{\Phi} } = \fracpart{ \theta - \Phi } $ and so, the phase noise may equivalently be modelled as normally distributed \emph{without wrapping} ~\cite{Hassibi98,WenchaoLi2013}.
For all integers $k$, 
\begin{equation}
Y = \fracpart{ r_0/\lambda + \Phi } = \sfracpart{(r_0 + k\lambda)/\lambda + \Phi},
\end{equation}
and so, the range is identifiable only if $r_0$ is assumed to lie in an interval of length $\lambda$.  A natural choice is the interval $[0, \lambda)$. This poses a problem if the range $r_0$ is larger than the wavelength $\lambda$.  To alleviate this, a common approach is to transmit multiple signals %$x_1(t),\ldots,x_N(t)$
$x_n(t) = e^{2\pi(f_nt + \phi)}$ for $n = 1,\dots,N$, each with a different frequency $f_n$.  Now $N$ phase estimates $\hat{\theta}_1,\dots,\hat{\theta}_N$ are computed along with phase differences 
\begin{equation}\label{eq:Yndefn}
Y_n = \sfracpart{\phi - \hat{\theta}_n} = \fracpart{ r_0/\lambda_n + \Phi_n} \qquad n = 1,\dots,N
\end{equation}
where $\lambda_n = c/f_n$ is the wavelength of the $n$th signal and $\Phi_1,\dots,\Phi_N$ represent phase noise.  Given $Y_1,\dots,Y_N$, a pragmatic estimator of the range $r_0$ is a minimiser of the least squares objective function
\begin{equation}
LS(r) = \sum_{n=1}^N \fracpart{Y_n - r/\lambda_n}^2.
\end{equation}
This least squares estimator is also the maximum likelihood estimator under the assumption that the phase noise variables $\Phi_1,\dots,\Phi_N$ are independent and identically wrapped normally distributed with zero mean~\cite[p.~50]{Mardia_directional_statistics}\cite[p.~76]{McKilliam2010thesis}\cite[p.~47]{Fisher1993}.

The objective function $LS$ is periodic with period equal to the smallest positive real number $P$ such that $P/\lambda_n \in \ints$ for all $n=1,\dots,N$, that is, $P = \lcm(\lambda_1,\dots,\lambda_N)$ is the least common multiple of the wavelengths.  The range is identifiable if we assume $r_0$ to lie in an interval of length $P$.  A natural choice is the interval $[0,P)$ and we correspondingly define the least squares estimator of the range $r_0$ as
\begin{equation}\label{eq:hatdininterval}
\hat{r} = \arg\min_{r \in [0,P)} LS(r).
\end{equation}
If $\lambda_n/\lambda_k$ is irrational for some $n$ and $k$ then the period $P$ does not exist and the objective function $LS$ is not periodic.  In this paper we assume this is not the case and that a finite period $P$ does exist.

\section{Lattice theory}\label{sec:lattice-theory}

Let $\Bbf$ be the $m\times n$ matrix with linearly independent column vectors $\bbf_1,\dots,\bbf_n$ from $m$-dimensional Euclidean space $\reals^m$ with $m\geq n$. The set of vectors
\[
\Lambda=\{ \Bbf\ubf \mid \ubf \in \ints^n \}
\]
is called an $n$-dimensional \term{lattice}.
%Let $\mathbf{b}_1,....,\mathbf{b}_n$ be linearly independent vectors from $m$-dimensional Euclidean space $\reals^m$ with $m\geq n$.  The set of vectors
%\[
%\Lambda = \{ u_1\bbf_1 + \dots + u_n \bbf_n \mid u_1,\dots,u_n \in \ints \}
%\]
%is called an $n$-dimensional \term{lattice}.  The elements of $\Lambda$ are called \term{lattice points} or \term{lattice vectors}. 
%The vectors $\bbf_1,\dots,\bbf_n$ form a \emph{basis} for the lattice $\Lambda$.  We can equivalently write
%\[
%\Lambda=\{ \Bbf\ubf \mid \ubf \in \ints^n \}
%\]
%where $\Bbf$ is the $m\times n$ matrix with columns $\bbf_1,\dots,\bbf_n$.  
%Unless otherwise stated vectors are column vectors in this paper. 
The matrix $\Bbf$ is called a \term{basis} or \term{generator} for $\Lambda$.  The basis of a lattice is not unique. If $\Ubf$ is an $n \times n$ matrix with integer elements and determinant $\det\Ubf=\pm 1$ then  $\Ubf$ is called a \term{unimodular matrix} and $\Bbf$ and $\Bbf\Ubf$ are both bases for $\Lambda$.  %When $m = n$ the lattice is said to be \term{full rank}. When $m > n$ the lattice points lie in the $n$-dimensional subspace of $\reals^m$ spanned by $\bbf_1,\dots,\bbf_n$.  
The set of integers $\ints^n$ is called the \term{integer lattice} with the $n\times n$ identity matrix $\Ibf$ as a basis.
%The parallelepiped formed by basis vectors $\bbf_1,\dots,\bbf_n$ is called a \term{fundamental parallelepiped} of the lattice $\Lambda$.  A fundamental parallelepiped has $n$-dimensional volume $\sqrt{\det \Bbf^\prime\Bbf }$ where superscript $^\prime$ denotes the vector or matrix transpose.  This quantity is also called the determinant of the lattice and is denoted by $\det\Lambda$.  The (closed) \term{Voronoi cell}, denoted $\vor\Lambda$, of an $n$-dimensional lattice $\Lambda$ in $\reals^m$ is the subset of $\reals^m$ containing all points nearer or of equal distance (here with respect to the Euclidean norm) to the lattice point at the origin than to any other lattice point. The Voronoi cell is an $m$-dimensional convex polytope that is symmetric about the origin.  If the lattice is full rank so that $n=m$ then the volume of the Voronoi cell is equal to the volume of a fundamental parallelepiped, that is, $\det\Lambda$.  Otherwise, if $m > n$ the Voronoi cell is unbounded in those directions orthogonal to the subspace spanned by the basis vectors $\bbf_1,\dots,\bbf_n$.  In this case the intersection of the Voronoi cell with this subspace has $n$-dimensional volume equal to $\det\Lambda$.
Given a lattice $\Lambda$ its \term{dual lattice}, denoted $\Lambda^*$, contains those points that have integral inner product with all points from $\Lambda$, that is,
\[
\Lambda^* = \{ \xbf  \mid \xbf^\prime \ybf \in \ints \text{ for all } \ybf \in \Lambda \}.
\]
The following proposition follows as a special case of Proposition~1.3.4 and Corollary~1.3.5 of \cite{Martinet2003}.

\begin{proposition} \label{cor:intlatticedim1}
Let $\vbf\in\ints^n$, let $H$ be the $n-1$ dimensional subspace orthogonal to $\vbf$, and let
\[
\Qbf = \Ibf - \frac{\vbf\vbf^\prime}{\vbf^\prime\vbf} = \Ibf - \frac{\vbf\vbf^\prime}{\|\vbf\|^2}
\]
be the $n\times n$ orthogonal projection matrix onto $H$.  The set of vectors $\ints^n\cap H$ is an $n-1$ dimensional lattice with % determinant 
% \[
% \det(\ints^n \cap H) = \|\vbf\|
% \]
% and 
dual lattice $(\ints^n \cap H)^* = \{ \Qbf \zbf \mid \zbf \in \ints^n \}$.
\end{proposition}

Given a lattice $\Lambda$ in $\reals^m$ and a vector $\ybf \in \reals^m$, a problem of interest is to find a lattice point $\xbf \in \Lambda$ such that the squared Euclidean norm $\| \ybf - \xbf \|^2 = \sum_{i=1}^m (y_i - x_i)^2$ is minimised.  This is called the \term{closest lattice point problem} (or \term{closest vector problem}) and a solution is called a \term{closest lattice point} (or simply \term{closest point}) to $\ybf$ \cite{Agrell2002}.  %The closest lattice point problem and the Voronoi cell are related in that $\xbf\in\Lambda$ is a closest lattice point to $\ybf$ if and only $\ybf - \xbf \in \vor(\Lambda)$.  

The closest lattice point problem is known to be NP-hard~\cite{micciancio_hardness_2001, Jalden2005_sphere_decoding_complexity}. Nevertheless, algorithms exist that can compute a closest lattice point in reasonable time if the dimension is small (less that about 60)~\cite{Kannan1987_fast_general_np,schnorr_euchner_sd_1994,Viterbo_sphere_decoder_1999,Agrell2002,MicciancioVoulgaris_deterministic_jv_2013}.  These algorithms have gone by the name ``sphere decoder'' in the communications engineering and signal processing literature. 
Although the problem is NP-hard in general, fast algorithms are known for specific highly regular lattices~\cite{McKilliam2009CoxeterLattices,McKilliam_closest_point_lattice_first_kind_2014}. For the purpose of range estimation the dimension of the lattice will be $N-1$ where $N$ is the number of frequencies transmitted.  The number of frequencies is usually small (less than 10) and, in this case, general purpose algorithms for computing a closest lattice point are fast~\cite{Agrell2002}. %We will not consider $N$ larger than TODO here.  In this case general purpose algorithms for computing a closest lattice point, such as those described in~\cite{Agrell2002}, are fast.  %For example, computing a closest point in a lattice of dimension TODO requires only a few milliseconds on a modern computer (see Figure~\ref{fig:sdbenchmarks}).
%\begin{figure}[t] 
%	\centering      
%	%	\includegraphics{figs/latticefigures-1.mps} 
%		\caption{TODO: Potentially add a figure showing benchmarks with the Schnor-Euchner sphere decoder and Babai's algorithm.}     
%		\label{fig:sdbenchmarks}   
%\end{figure} 

%**************************************************************************************************************************************************************
%		Section IV :  Range estimation and the closest lattice point problem
%**************************************************************************************************************************************************************
\section{Range estimation and the closest lattice point problem} \label{sec:range-estim-clos}

In this section we show how the least squares range estimator $\hat{r}$ from~\eqref{eq:hatdininterval} can be efficiently computed by computing a closest point in a lattice of dimension $N-1$.  The derivation is similar to those in~\cite{McKilliamFrequencyEstimationByPhaseUnwrapping2009,McKilliam_mean_dir_est_sq_arc_length2010,McKilliam_pps_unwrapping_tsp_2014}.  Our notation will be simplified by the change of variable $r = P\beta$, where $P$ is the least common multiple of the wavelengths.  Put $v_n = P/\lambda_n \in \ints$ and define the function
\[
F(\beta) = LS(P\beta) = \sum_{n=1}^N\fracpart{Y_n - \beta v_n}^2.
\]
Because $LS$ has period $P$ it follows that $F$ has period $1$.  If $\hat{\beta}$ minimises $F$ then $P\hat{\beta}$ minimises $LS$ and, because $\hat{r} \in [0,P)$, we have $\hat{r} = P( \hat{\beta} - \sfloor{\hat{\beta}} )$.  It is thus sufficient to find a minimiser $\hat{\beta} \in \reals$ of $F$.

Observe that $\fracpart{Y_n - \beta v_n}^2 = \min_{z \in \ints} (Y_n - \beta v_n - z)^2$ and so $F$ may equivalently be written
\[
F(\beta) = \min_{z_1,\dots,z_N \in \ints} \sum_{n=1}^N (Y_n - \beta v_n - z_n)^2.
\]
The integers $z_1,\dots,z_N$ are often called \emph{wrapping variables} and are related to the number of whole wavelengths that occur over the range $r_0$ between transmitter and receiver. The minimiser $\hat{\beta}$ can be found by jointly minimising the function
\[
F_1(\beta, z_1,\dots,z_N) = \sum_{n=1}^N (Y_n - \beta v_n - z_n)^2
\] 
over the real number $\beta$ and integers $z_1,\dots,z_N$.  This minimisation problem can be solved by computing a closest point in a lattice.  %It is easier to see this if we write in vector form.  
To see this, define column vectors 
\begin{align*}
\ybf &= (Y_1,\dots,Y_N)^\prime \in \reals^N, \\
\zbf &= (z_1,\dots,z_N)^\prime \in \ints^N, \\
\vbf &= (v_1,\dots,v_N)^\prime = \left(P/\lambda_1,\dots,P/\lambda_N\right)^\prime \in \ints^N.
\end{align*}
Now
\[
F_1(\beta, z_1,\dots,z_N) = F_1(\beta, \zbf) =  \| \ybf - \beta\vbf - \zbf \|^2.
\]
The minimiser of $F_1$ with respect to $\beta$ as a function of $\zbf$ is
\[ 
\hat{\beta}(\zbf) = \frac{(\ybf - \zbf)^\prime\vbf}{\vbf^\prime\vbf}.
\] 
Substituting this into $F_1$ gives
\[
F_2(\zbf) = \min_{\beta \in \reals} F_1(\beta, \zbf) = F_1\big(\hat{\beta}(\zbf), \zbf\big) = \| \Qbf\ybf - \Qbf\zbf \|^2
\]
where $\Qbf = \Ibf - \vbf\vbf^\prime/\|\vbf\|^2$ is the orthogonal projection matrix onto the $N-1$ dimensional subspace orthogonal to $\vbf$.  Denote this subspace by $H$.  By Proposition~\ref{cor:intlatticedim1} the set $\Lambda = \ints^N \cap H$ is an $N-1$ dimensional lattice with dual lattice $\Lambda^* = \{ \Qbf \zbf \mid \zbf \in \ints^N \}$.  We see that the problem of minimising $F_2(\zbf)$ is precisely that of finding a closest point in the lattice $\Lambda^*$ to $\Qbf\ybf \in \reals^N$.  Suppose we find $\hat{\xbf} \in \Lambda^*$ closest to $\Qbf \ybf$ and a corresponding $\hat{\zbf} \in \ints^N$ such that $\hat{\xbf} = \Qbf\hat{\zbf}$.  Then $\hat{\zbf}$ minimises $F_2$ and $\hat{\beta}(\hat{\zbf})$ minimises $F$.  The least squares range estimator in the interval $[0,P)$ is then
\begin{equation}\label{eq:leastsquaresrangehatz}
\hat{r} = P\big(\hat{\beta}(\hat{\zbf}) - \sfloor{\hat{\beta}(\hat{\zbf})}\big). %= P\left(\frac{(\ybf - \hat{\zbf})^\prime\vbf}{\vbf^\prime\vbf} - \floor{\frac{(\ybf - \hat{\zbf})^\prime\vbf}{\vbf^\prime\vbf}} \right)
\end{equation}

It remains to provide a method to compute a closest point $\hat{\xbf} \in \Lambda^*$ and a corresponding $\hat{\zbf} \in \ints^N$.  In order to use known general purpose algorithms we must first provide a basis for the lattice $\Lambda^*$~\cite{Agrell2002}.  The projection matrix $\Qbf$ is \emph{not} a basis because it is not full rank.  As noted by Li~et.~al.~\cite{Li_distance_est_wrapped_phase}, a modification of the Lenstra-Lenstra-Lovas algorithm due to Pohst~\cite{Pohst_modified_LLL_reduced_rank_1987} can be used to compute a basis given $\Qbf$.  However, it is preferable to have an explicit construction and Li~et.~al.~\cite{Li_distance_est_wrapped_phase} give a construction under the assumption that the wavelengths $\lambda_1,\dots,\lambda_N$ can be scaled to relatively prime integers, that is, there exists $c \in \reals$ such that $\gcd(c\lambda_k,c\lambda_n) = 1$ for all $k \neq n$.\footnote{The assumption that the wavelengths are pairwise relatively prime is made implicitly in equation (75) in~\cite{Li_distance_est_wrapped_phase}.}  We now remove the need for this assumption and construct a basis in the general case.  As a secondary benefit, we believe our construction to be simpler than that in~\cite{Li_distance_est_wrapped_phase}.  The following proposition is required. 

\begin{proposition}\label{prop:unimodMbasisfinder}
Let $\mathbf{U}$ be an $N \times N$ unimodular matrix with first column given by $\vbf$.  A basis for the lattice $\Lambda^*$ is given by the projection of the last $N-1$ columns of $\mathbf{U}$ orthogonally onto $H$.  That is, $\Qbf\ubf_2, \dots, \Qbf\ubf_{N}$ is a basis for $\Lambda^*$ where $\ubf_1,\dots,\ubf_{N}$ are the columns of $\Ubf$.
\end{proposition}
\begin{IEEEproof}
Because $\mathbf{U}$ is unimodular it is a basis matrix for the integer lattice $\ints^N$.  So, every lattice point $\zbf \in \ints^N$ can be uniquely written as $\zbf = c_1 \ubf_1 + \dots + c_{N} \ubf_{N}$ where $c_1,\dots,c_N\in\ints$.  The lattice 
\begin{align*}
\Lambda^* &= \{ \Qbf\zbf \mid \zbf \in \ints^N \} \\
&= \{ \Qbf (c_1 \ubf_1 + \dots + c_{N} \ubf_{N})  \mid c_1,\dots,c_N\in\ints \} \\
&= \{ c_2 \Qbf\ubf_2 + \dots + c_{N} \Qbf\ubf_{N}  \mid c_{2},\dots,c_{N}\in\ints \}
\end{align*}
because $\Qbf\mathbf{u}_1 = \Qbf\vbf = \zerobf$ is the origin.  It follows that $\Qbf\ubf_2,\dots,\Qbf\ubf_{N}$ form a basis for $\Lambda^*$.
\end{IEEEproof}

To find a basis for $\Lambda^*$ we require a matrix $\mathbf{U}$ as described by the previous proposition.  Such a matrix is given by Li~et.~al.~\cite[Eq.~(76)]{Li_distance_est_wrapped_phase} under the assumption that the wavelengths can be scaled to pairwise relatively prime integers.  We do not require this assumption here.  %This is important because the optimised frequencies for range estimation may not be pairwise relatively prime. In the following, we describe that how to find the matrix $\mathbf{U}$ under the general case when the wavelengths $\lambda_1,\dots,\lambda_N$ are not pairwise relatively prime integers. The proposed method is valid for any real number wavelengths until they share a least common multiple.
Because $P = \lcm(\lambda_1,\dots,\lambda_N)$ it follows that the integers $v_1,\dots,v_N$ are \emph{jointly} relatively prime, that is, $\gcd(v_1,\dots,v_N) = 1$.  Define integers $g_{1},\dots,g_N$ by $g_N = v_N$ and
\[
g_k = \gcd(v_k,\dots,v_N) = \gcd(v_k,g_{k+1}), \;\; k = 1,\dots,N-1
\]
and observe that $g_{k+1}/g_k$ and $v_k/g_k$ are relatively prime integers.  For $k = 1,\dots,N-1$, define the $N$ by $N$ matrix $\Abf_k$ with $m,n$th element
\[
A_{kmn} = \begin{cases}
v_k/g_k & m=n=k \\
g_{k+1}/g_k & m=k+1, n=k \\
a_k & m=k, n=k+1 \\
b_k & m=n=k+1 \\
I_{mn} & \text{otherwise}
\end{cases}
\]
where $I_{mn} = 1$ if $m=n$ and $0$ otherwise.  The integers $a_k$ and $b_k$ are chosen to satisfy 
\begin{equation}\label{eq:akbkexteuc}
b_k \frac{v_k}{g_k} - a_k \frac{g_{k+1}}{g_k} = 1
\end{equation}
and can be computed by the extended Euclidean algorithm.  The matrix $\Abf_k$ is equal to the identity matrix everywhere except at the $2$ by $2$ block of indices $k \leq m \leq k+1$ and $k \leq n \leq k+1$.  % For example,
% \[
% \Abf_{1} =  \left(\begin{array}{ccccc}
% v_1 & a_{1} & 0 & \cdots & 0 \\
% g_2 & b_{1} & 0 & \cdots & 0 \\
% 0 & 0 & 1 & \cdots & 0 \\
% \vdots & \vdots & \vdots & \ddots & \vdots \\
% 0 & 0 & 0 & \cdots & 1 
% \end{array}
% \right)
% \]
% because $g_1 = \gcd(v_1,\dots,v_N) = 1$.
The matrix $\Abf_k$ is unimodular for each $k$ because it has integer elements and because the determinant of the $2$ by $2$ matrix
\[
\left\vert\begin{array}{cc}
v_k/g_k & a_{k} \\
g_{k+1}/g_k & b_{k}  
\end{array}
\right\vert = b_k \frac{v_k}{g_k} - a_k \frac{g_{k+1}}{g_k} = 1
\]
as a result of~\eqref{eq:akbkexteuc}. A matrix $\Ubf$ satisfying the requirements of Proposition~\ref{prop:unimodMbasisfinder} is now given by the product
\[
\Ubf = \prod_{k=1}^{N-1} \Abf_k = \Abf_{N-1}\times \Abf_{N-2} \times \dots \times \Abf_1.
\]
That $\Ubf$ is unimodular follows immediately from the unimodularity of $\Abf_1,\dots,\Abf_{N-1}$.  It remains to show that the first column of $\Ubf$ is equal to $\vbf$.  Let $\vbf_1,\dots,\vbf_{N-1}$ be column vectors of length $N$ defined as
\begin{align*}
&\vbf_k = (v_1,\dots,v_k, g_{k+1},0,\dots,0)^\prime, \qquad k = 1,\dots,N-2 \\
&\vbf_{N-1} = (v_1,\dots,v_{N-1}, g_{N})^\prime = \vbf.
\end{align*}
One can readily check that $\vbf_{k+1} = \Abf_{k+1}\vbf_k$ for all $k=1,\dots,N-1$.  The first column of the matrix $\Abf_1$ is $\vbf_1$ and so, by induction, the first column of the product $\prod_{k=1}^K \Abf_k$ is $\vbf_K$ for all $K = 1,\dots,N-1$. It follows that the first column of $\Ubf$ is $\vbf_{N-1} = \vbf$ as required.  %We have asserted that $\Ubf$ satisfies the requirements of Proposition~\ref{prop:unimodMbasisfinder}.

Let $\Ubf_2$ be the $N$ by $N-1$ matrix formed by removing the first column from $\Ubf$, that is, $\Ubf_2 =(\ubf_2,\dots,\ubf_N)$.  By Proposition~\ref{prop:unimodMbasisfinder} a basis for $\Lambda^*$ is given by projecting the columns of $\Ubf_2$ orthogonally to $\vbf$, that is, a basis matrix for $\Lambda^*$ is the $N$ by $N-1$ matrix $\Bbf = \Qbf\Ubf_2$.  Given $\Bbf$ a general purpose algorithm~\cite{Agrell2002} can be used to compute $\hat{\wbf} \in \ints^{N-1}$ such that $\hat{\xbf} = \Bbf\hat{\wbf}$ is a closest lattice point in $\Lambda^*$ to $\Qbf\ybf \in \reals^{N}$.  Now
\[
\hat{\xbf} = \Bbf\hat{\wbf} = \Qbf\Ubf_2\hat{\wbf} = \Qbf\hat{\zbf}
\]
and so $\hat{\zbf} = \Ubf_2\hat{\wbf} \in \ints^N$.  The least squares range estimator $\hat{r}$ is then given by~\eqref{eq:leastsquaresrangehatz}. 

% Explicitly the basis vectors are
% \begin{align*}
% \bbf_{1} &= \Qbf(a_{12}\ebf_{1} + a_{22}\ebf_{2}) = a_{12}\ebf_{1} + a_{22}\ebf_{2} - \frac{a_{12}v_1 + a_{22}v_2}{\|\vbf\|^2}\vbf \\
% \bbf_{n} &= \Qbf\ebf_{n+1} = \ebf_{n+1} - \frac{v_{n+1}}{\|\vbf\|^2}\vbf \qquad n = 2,\dots,N-1
% \end{align*}
% where $\ebf_{n}$ denotes the vector with a one in the $n$th position and zeros elsewhere.

% The elements of $\Ubf$ are
% \[
% U_{mn} = \begin{cases}
% v_m & n = 1 \\ 
% a_{11} + a_{12}v_2 & n=m=1 \\
% a_{21} + a_{22}v_2 & m=2,n=1 \\
% a_{12} & m=1,n=2 \\
% a_{22} & m=n=2 \\
% 1 & n=m>1 \\
% 0 & \text{otherwise}
% \end{cases} 
% \]
%\newcommand*{\Resize}[2]{\resizebox{#1}{!}{$#2$}}

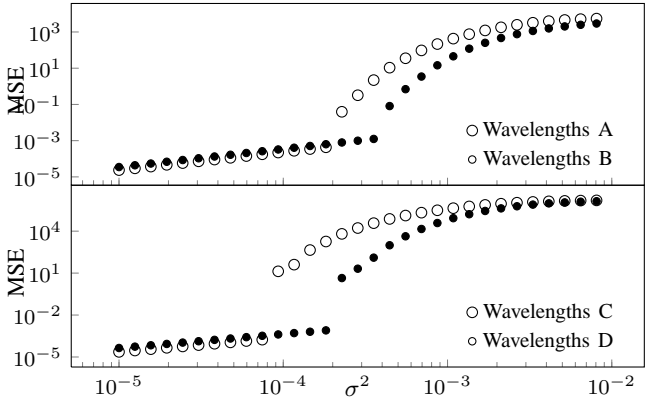
\begin{figure}[t] 
  \centering 
  \begin{tikzpicture} 
    \selectcolormodel{gray}
    \begin{groupplot}[
      group style={ 
        group name=my plots,
        group size=1 by 2,
        % ylabels at=edge left,
        xlabels at=edge bottom,
        xticklabels at=edge bottom,
        vertical sep=0pt
      },
      legend style={
         draw=none,
         fill=none,
         legend pos=south east,
       %  at={(0.75,1)},
       %  cells={anchor=west},
         font=\footnotesize
      },
      ylabel={MSE},
      ylabel style={at={(0.065,0.52)}},
      footnotesize,
      width=9.2cm,
      height=4cm,
      tickpos=left,
      ytick align=inside, 
      xtick align=inside,
      xmode=log, 
      ymode=log,
      ]

\nextgroupplot[]

\addplot[mark=o,only marks,mark options={scale=1}] table {code/data/LeastSquaresA};
% \addplot[mark=*,only marks,mark options={scale=0.5}] table {code/data/Babai};
\addplot[mark=*,only marks,mark options={scale=0.7}] table {code/data/LeastSquaresB};
% \addplot[mark=square*,only marks,mark options={scale=0.5}] table {code/data/BabaiMod};
%\addplot[mark= ] table {code/data/CRTA};
%\addplot[mark= ,dashed] table {code/data/CRTB};
%\addplot[mark= ] table {code/data/TwoStageCRTA};
%\addplot[mark=x,,only marks] table {code/data/TwoStageCRTB};
%\legend{Least Sq. A, Least Sq. B, CRT A, CRT B, 2-Stage CRT B} 
\legend{Wavelengths A, Wavelengths B} 
 
\nextgroupplot[
xlabel={$\sigma^2$}, 
xlabel style={at={(0.5,0.22)}},
ytick={10000,10,0.01,0.00001},
yticklabels={$10^4$,$10^1$,$10^{-2}$,$10^{-5}$}
]

\addplot[mark=o,only marks,mark options={scale=1}] table {code/data/LeastSquaresC};
% \addplot[mark=*,only marks,mark options={scale=0.5}] table {code/data/Babai};
\addplot[mark=*,only marks,mark options={scale=0.7}] table {code/data/LeastSquaresD};
% \addplot[mark=square*,only marks,mark options={scale=0.5}] table {code/data/BabaiMod};
%\addplot[mark= ] table {code/data/CRTC};
%\addplot[mark= ,dashed] table {code/data/CRTD};
%\addplot[mark= ] table {code/data/TwoStageCRTC}; 
%\addplot[mark= ,dashed] table {code/data/TwoStageCRTD};
%\legend{Least Sq. C, Least Sq. D, CRT C, CRT D}
\legend{Wavelengths C, Wavelengths D} 
 
\end{groupplot}
 
\end{tikzpicture}
  \caption{Comparison of the least squares range estimator with wavelengths $A$ (top) and $C$ (bottom) suitable for the basis in~\cite{Li_distance_est_wrapped_phase} and $B$ (top) and $D$ (bottom) suitable only for the basis described in this paper.  Sets $B$ and $D$ result in smaller mean square error when the noise variance $\sigma^2$ is greater than approximately $1.2\times 10^{-4}$ and $7\times 10^{-5}$ respectively.}\label{fig:Comp_with_Li1}
\end{figure}

\section{Simulation Results}\label{sec:simulation-results}

We present the results of Monte-Carlo simulations with the least squares range estimator.  Simulations with $N=4$ and $N=5$ wavelengths are performed.  For each case we consider two different set of wavelengths.  The first set is suitable for the basis of Li~et.~al.~\cite{Li_distance_est_wrapped_phase} and was used in the simulations in~\cite{Li_distance_est_wrapped_phase}.  The second set is suitable only for our basis.  In each simulation the true range $r_0 = 20$ and the phase noise variables $\Phi_1,\dots,\Phi_N$ are wrapped normally distributed, that is, $\Phi_n = \fracpart{X_n}$ where $X_1,\dots,X_N$ are independent and normally distributed with zero mean and variance $\sigma^2$.  In this case, the least squares estimator is also the maximum likelihood estimator.  Figure ~\ref{fig:Comp_with_Li1} shows the sample mean square error for $\sigma^2$ in the range $10^{-5}$ to $10^{-2}$ and $10^7$ Monte-Carlo trials used for each value of $\sigma^2$.  %computed exactly using the algorithm of Schnorr and Euchner~\cite{Agrell2002,schnorr_euchner_sd_1994}. 

For $N=4$ the two sets of wavelengths are 
\[
A = \{2, 3, 5, 7\}, \;\;\; B = \{\tfrac{210}{79}, \tfrac{210}{61}, \tfrac{210}{41}, \tfrac{210}{31}\}.
\]
For both sets the wavelengths are contained in the interval $[2,7]$ and $P = 210 = \lcm(A) = \lcm(B)$ so that the identifiable range is the same.  The wavelengths $A$ are relatively prime integers and are suitable for the basis of Li~et.~al.~\cite{Li_distance_est_wrapped_phase} and are used in the simulations in~\cite{Li_distance_est_wrapped_phase}.  The wavelengths $B$ are not suitable for the basis of~\cite{Li_distance_est_wrapped_phase} because they can not be scaled to pairwise relatively prime integers.  To see this, observe that the smallest positive number by which we can multiply the elements of $B$ to obtain integers is $c = \tfrac{6124949}{210}$.  Multiplying the elements of $B$ by $c$ we obtain the set
\[
c \times B = \{77531, 100409, 149389, 197579 \}
\]  
and these elements are not pairwise relatively prime because, for example, $\gcd(77531, 100409) = 1271$.  Figure~\ref{fig:Comp_with_Li1} shows the results of simulations with both sets $A$ and $B$.  When the noise variance $\sigma^2$ is small wavelengths $A$ result in slightly reduced sample mean square error as compared with $B$.  As $\sigma^2$ increases the sample mean square error exhibits a `threshold' effect and increases suddenly.  The threshold occurs at $\sigma^2 \approx 1.2\times 10^{-4}$ for wavelengths $A$ and $\sigma^2 \approx 3\times 10^{-4}$ for wavelength $B$.  Wavelengths $B$ are more accurate than $A$ when $\sigma^2$ is greater than approximately $1.2\times 10^{-4}$.

For $N=5$ the two sets of wavelengths are
\[
C = \{2, 3, 5, 7, 11\}, \;\;\; D = \{\tfrac{2310}{877}, \tfrac{2310}{523}, \tfrac{2310}{277}, \tfrac{2310}{221}, \tfrac{2310}{211}\}.
\]
For both sets all wavelengths are contained in the interval $[2,11]$ and $P = 2310 = \lcm(C) = \lcm(D)$ so that the maximum identifiable range is the same. The basis of Li~et.~al.~\cite{Li_distance_est_wrapped_phase} can be used for wavelengths $C$ but not for $D$.  The wavelengths $C$ were used in the simulations in~\cite{Li_distance_est_wrapped_phase}.  Figure~\ref{fig:Comp_with_Li1} shows the result of Monte-Carlo simulations with these wavelengths.  Wavelengths $C$ result in slightly smaller sampler mean square error than $D$ when $\sigma^2$ is small, but dramatically more error for $\sigma^2$ above the threshold occurring at $\sigma^2 \approx 7 \times 10^{-5}$.

The sets $B$ and $D$ have been selected based on a heuristic optimisation criterion.  The properties of this criterion are not yet fully understood and will be the subject of a future paper.

 \section{Conclusion}

We have considered least squares/maximum likelihood estimation of range from observation of phase at multiple wavelengths.  The estimator can be computed by finding a closest point in a lattice.  This requires a basis for the lattice.  Bases have previously been constructed under the assumption that the wavelengths can be scaled to relatively prime integers.  In this paper, we gave a construction in the general case and indicated by simulation that this can dramatically improve range estimates.  An open problem is how to select wavelengths to maximise the accuracy of the least squares estimator.  We will study this problem in future research.

{
\small
\bibliography{bib}
}

\end{document}